\documentclass[superscriptaddress,prc,aps,twocolumn,showpacs]{revtex4}
\usepackage{amsmath,amssymb}
\usepackage{graphicx}
\newcommand{\GWU}[1]
{\affiliation{The George Washington University, Washington, DC 20052-0001, USA}}
\newcommand{\Mainz}[1]
{\affiliation{Institut f\"ur Kernphysik, University of Mainz, D-55099 Mainz,Germany}}
\newcommand{\UCLA}[1]
{\affiliation{University of California Los Angeles, Los Angeles, California 90095-1547, USA}}
\newcommand{\Gatchina}[1]
{\affiliation{Petersburg Nuclear Physics Institute, 188300 Gatchina, Russia}}
\newcommand{\Glasgow}[1]
{\affiliation{SUPA School of Physics and Astronomy, University of Glasgow,
 Glasgow G12 8QQ, United Kingdom}}
\newcommand{\Kent}[1]
{\affiliation{Kent State University, Kent, Ohio 44242-0001, USA}}
\newcommand{\Bonn}[1]
{\affiliation{Helmholtz-Institut f\"ur Strahlen- und Kernphysik, University of Bonn,
 D-53115 Bonn, Germany}}
\newcommand{\Giessen}[1]
{\affiliation{II Physikalisches Institut, University of Giessen, D-3539 Giessen, Germany}}
\newcommand{\Pavia}[1]
{\affiliation{INFN Sezione di Pavia, I-27100 Pavia, Italy}}
\newcommand{\LPI}[1]
{\affiliation{Lebedev Physical Institute, 119991 Moscow, Russia}}
\newcommand{\Dalhousie}[1]
{\affiliation{Dalhousie University, Halifax, Nova Scotia B3H 4R2, Canada}}
\newcommand{\Halifax}[1]
{\affiliation{Saint Mary’s University, Halifax, Nova Scotia B3H 3C3, Canada}}
\newcommand{\Edinburgh}[1]
{\affiliation{School of Physics, University of Edinburgh, Edinburgh EH9 3JZ,
 United Kingdom}}
\newcommand{\Sackville}[1]
{\affiliation{Mount Allison University, Sackville, New Brunswick E4L 1E6, Canada}}
\newcommand{\Basel}[1]
{\affiliation{Institut f\"ur Physik, University of Basel, CH-4056 Basel, Switzerland}}
\newcommand{\INR}[1]
{\affiliation{Institute for Nuclear Research, 125047 Moscow, Russia}}
\newcommand{\Zagreb}[1]
{\affiliation{Rudjer Boskovic Institute, HR-10000 Zagreb, Croatia}}

\begin{document}

\title{Photoproduction of the $\omega$ meson on the proton near threshold}

\author{I.~I.~Strakovsky}\thanks{corresponding author, e-mail: igor@gwu.edu}\GWU \\
\author{S.~Prakhov}\thanks{corresponding author, e-mail: prakhov@ucla.edu}\GWU \\ \Mainz \\ \UCLA \\
\author{Ya.~I.~Azimov}\Gatchina \\
\author{P.~Aguar-Bartolom\'e}\Mainz \\
\author{J.~R.~M.~Annand}\Glasgow \\
\author{H.~J.~Arends}\Mainz \\
\author{K.~Bantawa}\Kent \\
\author{R.~Beck}\Bonn \\
\author{V.~Bekrenev}\Gatchina \\
\author{H.~Bergh\"auser}\Giessen \\
\author{A.~Braghieri}\Pavia \\
\author{W.~J.~Briscoe}\GWU \\
\author{J.~Brudvik}\UCLA \\
\author{S.~Cherepnya}\LPI \\
\author{R.~F.~B.~Codling}\Glasgow \\
\author{C.~Collicott}\Dalhousie \\ \Halifax \\
\author{S.~Costanza}\Pavia \\
\author{B.~T.~Demissie}\GWU \\
\author{E.~J.~Downie}\Mainz \\ \GWU \\
\author{P.~Drexler}\Giessen \\
\author{L.~V.~Fil'kov}\LPI \\
\author{D.~I.~Glazier}\Glasgow \\ \Edinburgh \\
\author{R.~Gregor}\Giessen \\
\author{D.~J.~Hamilton}\Glasgow \\
\author{E.~Heid}\GWU \\ \Mainz \\
\author{D.~Hornidge}\Sackville \\
\author{I.~Jaegle}\Basel \\
\author{O.~Jahn}\Mainz \\
\author{T.~C.~Jude}\Edinburgh \\
\author{V.~L.~Kashevarov}\Mainz \\ \LPI \\
\author{I.~Keshelashvili}\Basel\\
\author{R.~Kondratiev}\INR \\
\author{M.~Korolija}\Zagreb \\
\author{M.~Kotulla}\Giessen \\
\author{A.~Koulbardis}\Gatchina \\
\author{S.~Kruglov}\thanks{deceased}\Gatchina \\
\author{B.~Krusche}\Basel \\
\author{V.~Lisin}\LPI \\
\author{K.~Livingston}\Glasgow \\
\author{I.~J.~D.~MacGregor}\Glasgow \\
\author{Y.~Maghrbi}\Basel\\
\author{D.~M.~Manley}\Kent \\ 
\author{Z.~Marinides}\GWU \\
\author{J.~C.~McGeorge}\Glasgow \\
\author{E.~F.~McNicoll}\Glasgow \\
\author{D.~Mekterovic}\Zagreb \\
\author{V.~Metag}\Giessen \\
\author{D.~G.~Middleton}\Mainz \\ \Sackville \\
\author{A.~Mushkarenkov}\Pavia \\ 
\author{B.~M.~K.~Nefkens}\thanks{deceased}\UCLA \\
\author{A.~Nikolaev}\Bonn \\ 
\author{R.~Novotny}\Giessen \\
\author{H.~Ortega}\Mainz \\
\author{M.~Ostrick}\Mainz \\  
\author{P.~B.~Otte}\Mainz \\
\author{B.~Oussena}\GWU \\ \Mainz \\
\author{P.~Pedroni}\Pavia \\
\author{F.~Pheron}\Basel \\
\author{A.~Polonski}\INR \\  
\author{J.~Robinson}\Glasgow \\
\author{G.~Rosner}\Glasgow \\
\author{T.~Rostomyan}\Basel \\
\author{S.~Schumann}\Mainz \\
\author{M.~H.~Sikora}\GWU \\ \Edinburgh \\
\author{A.~Starostin}\UCLA \\
\author{I.~Supek}\Zagreb \\
\author{M.~F.~Taragin}\GWU \\
\author{C.~M.~Tarbert}\Edinburgh \\
\author{M.~Thiel}\Giessen \\
\author{A.~Thomas}\Mainz \\   
\author{M.~Unverzagt}\Mainz \\ \Bonn \\ 
\author{D.~P.~Watts}\Edinburgh \\
\author{D.~Werthm\"uller}\Basel \\
\author{F.~Zehr}\Basel \\

\collaboration{A2 Collaboration at MAMI}

\date{\today}

\begin{abstract}
An experimental study of $\omega$ photoproduction on the proton was
conducted by using the Crystal Ball and TAPS multiphoton spectrometers
together with the photon tagging facility at the Mainz Microtron MAMI.
The $\gamma p\to\omega p$ differential cross sections are measured from
threshold to the incident-photon energy $E_\gamma=1.40$~GeV
($W=1.87$~GeV for the center-of-mass energy) with 15-MeV binning
in $E_\gamma$ and full production-angle coverage.
The quality of the present data near threshold gives access
to a variety of interesting physics aspects. As an example,
an estimation of the $\omega N$ scattering length $\alpha_{\omega p}$
is provided.
\end{abstract}


\maketitle

\section{Introduction}

Although Quantum Chromodynamics (QCD) is generally believed to govern
strong interactions, it still can be applied to particular problems only
in terms of specific model-dependent approaches, which can be distinguished
by their predictions for the resonance spectra.
 However, such predictions can only be verified
 in collisions of a very restricted set of hadron pairs:
 a proton or a bound neutron as a target and
 a stable or a weakly decaying hadron as a projectile.
 Meanwhile, various constituent quark models
 ({\it e.g.}, see Ref.~\cite{CapRob} and references therein)
 predict a richer spectrum of hadron resonances
 than have so far been observed in experiments~\cite{PDG}.
 Because most known baryon states
 were discovered in elastic $\pi N$ scattering, some resonances could
 have been missed because of their weak coupling to the $\pi N$
 channels~\cite{ar06}. At the same time, a stronger coupling
 of those resonances to such channels
 as $\eta N$, $\omega N$, $K\Lambda$, or $K\Sigma$ cannot be excluded,
 and, therefore, an extensive study of these channels is very important
 in searching for the so-called ``missing'' resonances~\cite{Koniuk}.
 Proof of their existence would constitute a strong confirmation
 of the validity of the constituent quark concept.

Although the $\omega$ meson is neither stable nor decaying weakly,
the $\omega N$ channel is favorable in searching for
missing resonances because, owing to vector-meson dominance (VMD)~\cite{VMD},
$\omega$ photoproduction on the nucleon may be directly related
to the elastic amplitude for $\omega N$ scattering.
In addition, $\omega$ photoproduction provides an ``isospin filter" for
the nucleon response because $\omega N$ final states can originate
only from $N^*$ states with $I = 1/2$, but not from $\Delta^*$s
with $I = 3/2$.
 The $\omega N$ threshold region is also especially attractive
 in searching for new resonances because
 the reaction threshold is located at the higher-energy edge
 of the third resonance region, in which
 the Review of Particle Physics (RPP)~\cite{PDG} shows
 seven $N^*$ states with masses between 1650 and 1720~MeV,
 and then there are no observed $N^*$ states up to 1860~MeV.
 It cannot be excluded that this energy range may contain
 unknown $N^*$ resonances that are coupled more strongly to $\omega N$
 than to other meson-baryon channels. Such a strong coupling to
 resonances in the near-threshold region is clearly seen, for example,
 for the two most dominant channels, $\pi N$ and $\eta N$,
 coupled to $\Delta(1232)3/2^+$ and $N(1535)1/2^-$, respectively.

The photoproduction of $\omega$ mesons was under extensive
theoretical discussion since the first high-statistics differential
cross sections and polarization results were provided from
SAPHIR~\cite{SAPHIR}, covering a broad interval in center-of-mass (c.m.)
energy, up to $W=2.4$~GeV with $\sim 60$\% of the full production-angle
range. Many models, including coupled channel, effective
Lagrangian, and QCD-inspired
approaches~\cite{Beijing,Titov1,ba02,Titov2,Giessen,Paris}, were
developed, trying to explain the behavior of the experimental data.
Another high-statistics measurement of the $\gamma p\to\omega p$
differential cross sections and spin-density matrix elements
was recently made by the CLAS Collaboration~\cite{CLAS1},
covering c.m.\ energies from threshold up to $W=2.84$~GeV.
Their binning in c.m.\ energy was much finer than
from SAPHIR~\cite{SAPHIR}, but the production-angle coverage
was narrower. Unfortunately, for both SAPHIR and CLAS,
the data obtained in the energy region near threshold
have the poorest quality.

 Both SAPHIR~\cite{SAPHIR} and CLAS~\cite{CLAS1} experiments
 measured the photoproduction of $\omega$ via its major
 decay mode $\omega\to\pi^+\pi^-\pi^0$, having 89.2\% branching
 ratio~\cite{PDG}. The analyses of those experimental data were 
 challenging because of large background from
 $\gamma p\to \pi^+\pi^-\pi^0 p$ produced via intermediate
 states different from $\gamma p\to\omega p$.
 In the analysis by SAPHIR~\cite{SAPHIR}, all data points were
 obtained by individual fits of the $\omega$ peak above
 the $\pi^+\pi^-\pi^0$ background. Their fitting procedure
 experienced difficulties with describing the $\omega$ peak
 near threshold as a standard Breit-Wigner (BW) shape was
 severely distorted by phase space, and a function
 (BW with parameters of $\omega$ convoluted with two Gaussians
 describing the experimental resolutions) used for higher energies
 did not work here.
 In the analysis by CLAS~\cite{CLAS1}, a so-called $Q$-factor
 technique was used to separate signal and background events,
 allowing to fit data with the event-by-event approach and
 to extract all results from their unbinned maximum-likelihood fits.  
 
 A background contamination of $\omega$ events would be expected
 to improve if $\omega$ photoproduction were measured in process
 $\gamma p\to \omega p \to \pi^0\gamma p \to 3\gamma p$,
 using a radiative decay mode, $\omega\to\pi^0\gamma$,
 with 8.28\% branching ratio~\cite{PDG}. Other physical
 processes having the three-photon final state are expected
 to be very small. 
 However, as shown in a preliminary analysis of the CBELSA/TAPS
 data~\cite{CBELSA}, the three-photon background can come from
 processes $\gamma p\to \pi^0\pi^0 p\to 4\gamma p$ and
 $\gamma p\to \pi^0\eta p\to 4\gamma p$
 when one of the four final-state photons was not detected.

In this work, a new high-statistics measurement of the
$\gamma p\to\omega p$ differential cross sections near
threshold is presented. The results are based on
an analysis of $\sim 5\times 10^5$ $\omega$ mesons
detected via their $\pi^0\gamma$ decay mode.
All data were divided into 20 (15~MeV each) bins
in the incident-photon energy, $E_\gamma$, and 15 angular bins,
covering the full range of the $\omega$ production angle.
All results are obtained by individual fits of the $\omega$ peak
above background in each energy-angle bin.

A partial-wave analysis with extracting spin-density matrix elements from
the present data is in progress, by using the $Q$-factor technique.
As the latter analysis is being performed by a theoretical group outside the A2
Collaboration, their results will be published later on separately.

\section{Experimental setup}
\label{sec:Setup}

The process $\gamma p\to \omega p \to \pi^0\gamma p \to 3\gamma p$
was measured using the Crystal Ball (CB)~\cite{CB}
as a central spectrometer and TAPS~\cite{TAPS,TAPS2}
as a forward spectrometer. These detectors were
installed at the energy-tagged bremsstrahlung-photon beam
produced from the electron beam of the Mainz Microtron (MAMI)~\cite{MAMI,MAMIC}.
In the present experiment, bremsstrahlung photons, produced by the 1508-MeV
electrons in a 10-$\mu$m Cu radiator and collimated by a 4-mm-diameter Pb collimator,
were incident on a 5-cm-long liquid-hydrogen (LH$_2$) target located
in the center of the CB.
The energies of the incident photons were analyzed up to 1402~MeV
by detecting the postbremsstrahlung electrons in the Glasgow tagging
spectrometer (or tagger)~\cite{TAGGER,TAGGER1,TAGGER2}.
The uncertainty in the energy of the tagged photons is mainly determined
by the width of tagger focal-plane detectors and the energy of
the MAMI electron beam used in experiments. For the MAMI energy of 1508~MeV,
such an uncertainty was about $\pm 2$~MeV.

The CB detector is a sphere consisting of 672
optically isolated NaI(Tl) crystals, shaped as
truncated triangular pyramids, which point toward
the center of the sphere. The crystals are arranged in two
hemispheres that cover 93\% of $4\pi$ sr, sitting
outside a central spherical cavity with a radius of
25~cm, which is designed to hold the target and inner
detectors. In this experiment, TAPS was
arranged in a plane consisting of 384 BaF$_2$
counters of hexagonal cross section.
It was installed 1.5~m downstream of the CB center
and covered the full azimuthal range for polar angles
from $1^\circ$ to $20^\circ$.

The experimental trigger required the total energy deposited in the CB
to exceed $\sim$320~MeV and the number of so-called hardware clusters
in the CB (multiplicity trigger) to be two or larger.
In the trigger, a hardware cluster in the CB was a block of 16
adjacent crystals in which at least one crystal had an energy
deposit larger than 30 MeV. The TAPS information was not used
in the trigger of the present experiment.

More details on the experimental setup, its resolutions, and other conditions
during the period of the data taking (first half of 2007) are given
in Refs.~\cite{etamamic,slopemamic} and references therein.

\section{Data handling}
\label{sec:Data}

 The reaction $\gamma p\to \omega p$ was searched for in
 events identified as $\gamma p\to \pi^0\gamma p \to 3\gamma p$
 candidates, having three or four clusters reconstructed
 in the CB and TAPS together by software analysis.
 The cluster algorithm in software was optimized for finding
 a group of adjacent crystals in which the energy was deposited
 by a single-photon electromagnetic shower. This algorithm also
 works well for a proton cluster. The software threshold
 for the cluster energy was chosen to be 12 MeV.
 For the $\gamma p\to \pi^0\gamma p \to 3\gamma p$
 candidates, the three-cluster events were analyzed
 assuming that the final-state proton was not detected.
 This typically happens when the outgoing proton is stopped in the material
 of the downstream beam tunnel of the CB, or the proton scatters in the backward
 direction within the c.m. frame, resulting in such a low kinetic energy of the
 proton in the laboratory system that it is below the software cluster
 threshold. Thus, including three-cluster events in the analysis is
 vital for measuring $\gamma p\to \omega p$ differential cross sections
 at very forward production angles of $\omega$. The fraction of 
 $\gamma p\to \omega p\to \pi^0\gamma p \to 3\gamma p$ events without
 the detected proton varies from 2.7\% at the reaction threshold
 to 7.6\% at the highest energy of the present experiment.

 The selection of event candidates and the reconstruction of the reaction
 kinematics was based on the kinematic-fit technique.
 Details on the kinematic-fit parametrization of the detector
 information and resolutions are given in Ref.~\cite{slopemamic}.
 As discussed there, the information for the outgoing proton,
 if it is detected, is used in the kinematic fit without
 the proton kinetic energy, which has large uncertainties because
 of the material between the target and the crystals of the calorimeters.
 In addition, when energetic protons punch through the crystals, their kinetic
 energy cannot be determined from their energy deposit.
 For three-clusters events, all parameters of the outgoing proton
 are free variables of the kinematic fit. All three- and four-clusters events
 that satisfied the $\gamma p\to \pi^0\gamma p \to 3\gamma p$ hypothesis
 with a probability greater than 2\% were accepted for further analysis.
 The kinematic fitting for this hypothesis includes the four main constraints,
 which are based on the conservation of energy and three-momentum, and
 an additional constraint on the invariant mass of two outgoing photons
 to have the $\pi^0$-meson mass. The kinematic-fit output was used
 to reconstruct the kinematics of the outgoing particles.
 Small misidentification of the proton with a photon for the selected
 $\gamma p\to \omega p\to \pi^0\gamma p \to 3\gamma p$ events was 
 observed only in the three-clusters events and only for clusters
 reconstructed in TAPS. Such events were discarded based on
 the time of flight between a TAPS cluster and the CB signal
 with respect to the energy of the TAPS cluster.

 The determination of the experimental acceptance was based on
 a Monte Carlo (MC) simulation of $\gamma p\to \omega p\to \pi^0\gamma p$
 with an isotropic production-angular distribution and an uniform
 beam distribution generated from the reaction threshold up to the maximal
 experimental energy. To reproduce the energy dependence of the $\omega$
 resonance shape near production threshold, the reaction
 $\gamma p\to \pi^0\gamma p$ was generated first as phase space.
 Then, the invariant mass of the $\pi^0\gamma$ system, $m(\pi^0\gamma)$,
 was folded with the Breit-Wigner (BW) function, the parameters of which
 were taken for the $\omega$ meson ($m=782.65$~MeV and $\Gamma=8.49$~MeV)
 from the RPP~\cite{PDG}. This approach allowed one to properly reproduce
 the folding of the BW shape with phase space.  
 All MC events were propagated through a GEANT (3.21) simulation
 of the experimental setup. To reproduce resolutions of the experimental data,
 the GEANT output was subject to additional smearing, thus allowing
 both the simulated and experimental data to be analyzed in the same way.
 Matching the energy resolution between the experimental and MC events
 was adjusted via reaching agreement in the invariant-mass resolutions and
 in the kinematic-fit stretch functions (or pulls) and probability
 distributions. Such an adjustment was based on the analysis of the
 same data set for reactions having almost no physical background
 (namely, $\gamma p\to \pi^0 p$, $\gamma p\to \eta p\to \gamma\gamma p$,
 and $\gamma p\to \eta p\to 3\pi^0p$~\cite{slopemamic}).
 After taking into account the trigger requirements in the analysis
 of the MC events, the average acceptance for the process
 $\gamma p\to \omega p\to \pi^0\gamma p$ in the A2 experimental setup
 was found to be about 50\%. The simulation of the trigger was adjusted
 with the same reactions that was used for adjusting the energy resolution,
 where $\gamma p\to \pi^0 p$ and $\gamma p\to \eta p\to \gamma\gamma p$
 are especially sensitive to the multiplicity trigger, required two
 hardware clusters in the CB. The agreement of the $\gamma p\to \eta p$
 differential cross sections obtained from both the $\eta \to \gamma\gamma$  
 and the $\eta\to 3\pi^0$ decay modes was another cross-check for the correctness
 of the trigger simulation. As found out, the multiplicity trigger was not
 important for the present analysis, as all $\gamma p\to \omega p\to \pi^0\gamma p$
 events that passed the requirement on the CB total energy to exceed 320~MeV
 had more than one hardware cluster in the CB.
\begin{figure*}
\includegraphics[width=0.87\textwidth,bbllx=0.5cm,bblly=.35cm,bburx=20.cm,bbury=7.cm]{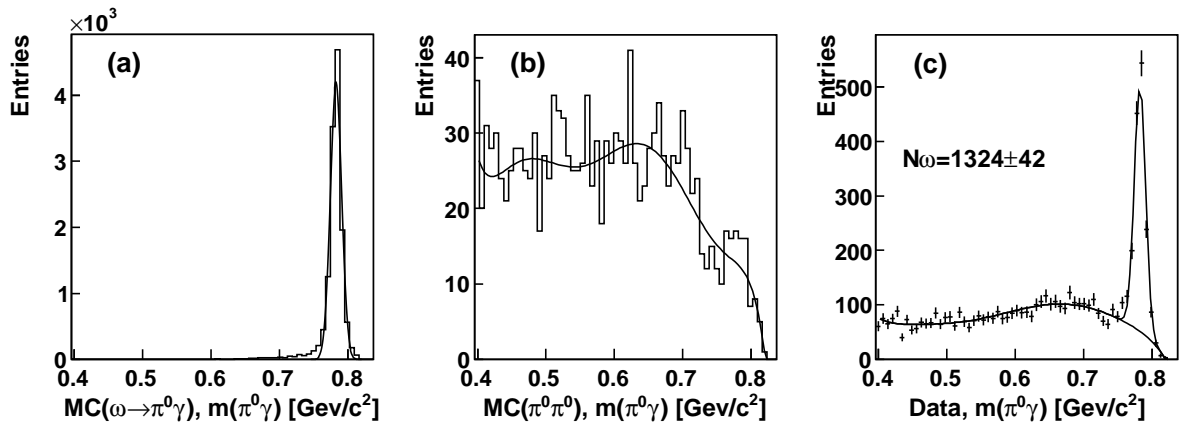}
\caption{
 $m(\pi^0\gamma)$ invariant-mass distributions obtained for $E_\gamma=1176$~MeV
 and $\cos\theta=0$.
 (a)~MC simulation of $\gamma p\to \omega p \to \pi^0\gamma p$ with a Gaussian fit
    ($\sigma=8.2$~MeV and FWHM~=~19.3~MeV).
 (b)~MC simulation of the background reaction $\gamma p\to \pi^0\pi^0p$
     fitted with a polynomial of order eight.
 (c)~Measured spectrum fitted with the sum of a Gaussian and a polynomial
    of order eight, where $N_\omega$ is the number of $\omega$ mesons
   found from this fit.
}
 \label{fig:mpi0g_fit1}
\end{figure*}

 As turned out, the selected experimental events with $\omega\to\pi^0\gamma$ decays
 were also contaminated with some background distributed quite smoothly under
 the $\omega$ peak. Based on the MC simulations of possible background reactions,
 it was found that processes $\gamma p\to \pi^0\pi^0 p\to 4\gamma p$ and
 $\gamma p\to \pi^0\eta p\to 4\gamma p$ could mimic $\gamma p\to\pi^0\gamma p$
 when one of the four final-state photons had not been detected.
 The background from these processes mostly contaminates the four-cluster events.
 Events from reaction $\gamma p\to \pi^0\pi^+ n$ can mimic $\gamma p\to\pi^0\gamma p$
 when the outgoing neutron was not detected, and they contaminate
 the three-cluster events. Nevertheless, there was no background process found
 that could mimic the $\omega\to\pi^0\gamma$ peak. 

 Because the suppression of the background processes was not possible without
 severe losses in $\omega\to\pi^0\gamma$ events themselves, and the subtraction
 of the background processes was not possible without precise MC simulations
 of all possible background reactions, the $\omega\to\pi^0\gamma$ events were
 measured by fitting experimental $m(\pi^0\gamma)$ spectra with some function,
 describing the $\omega$ peak above a smooth background.
 To measure the $\gamma p\to \omega p$ differential
 cross sections, all events were divided into 20 incident-photon energy
 bins from the reaction threshold to $E_\gamma=1402$~MeV.
 The data within each energy bin were divided
 into 15 identical $\cos\theta$ bins, covering the full range from -1 to 1,
 where $\theta$ was the angle between the directions of the outgoing
 $\pi^0\gamma$ system and the incident photon in the c.m.\ frame.
 The number of $\omega \to \pi^0\gamma$ decays observed
 in each energy-angle bin was determined by an individual fit
 of the corresponding $m(\pi^0\gamma)$ spectra.
\begin{figure*}
\includegraphics[width=0.87\textwidth,bbllx=0.5cm,bblly=.35cm,bburx=20.cm,bbury=7.cm]{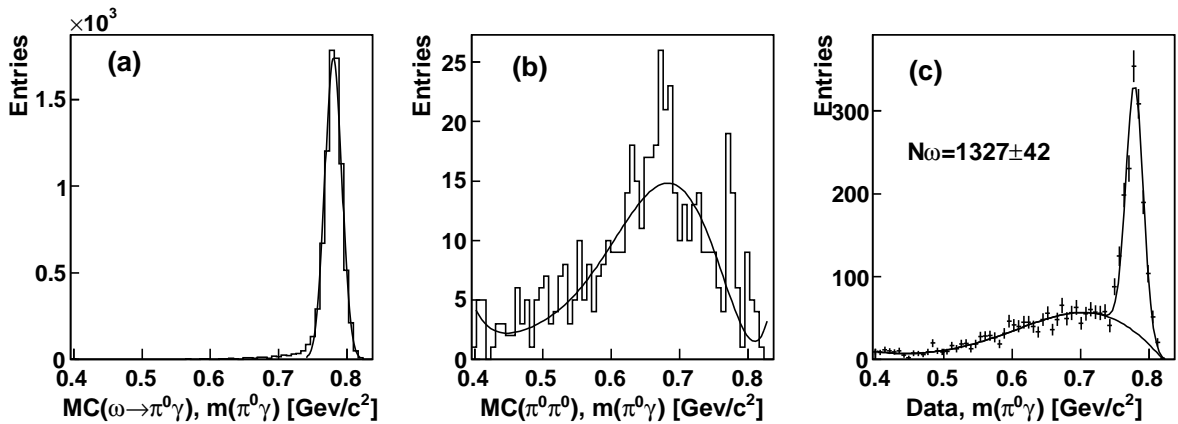}
\caption{
 Same as Fig.~\protect\ref{fig:mpi0g_fit1}, but for $\cos\theta=0.933$.
 A Gaussian fit to the $\omega$ peak results in $\sigma=12.4$~MeV (FWHM=29.2~MeV).
 A polynomial of order seven used to fit the background.
}
 \label{fig:mpi0g_fit2}
\end{figure*}
\begin{figure*}
\includegraphics[width=0.87\textwidth,bbllx=0.5cm,bblly=.35cm,bburx=20.cm,bbury=7.cm]{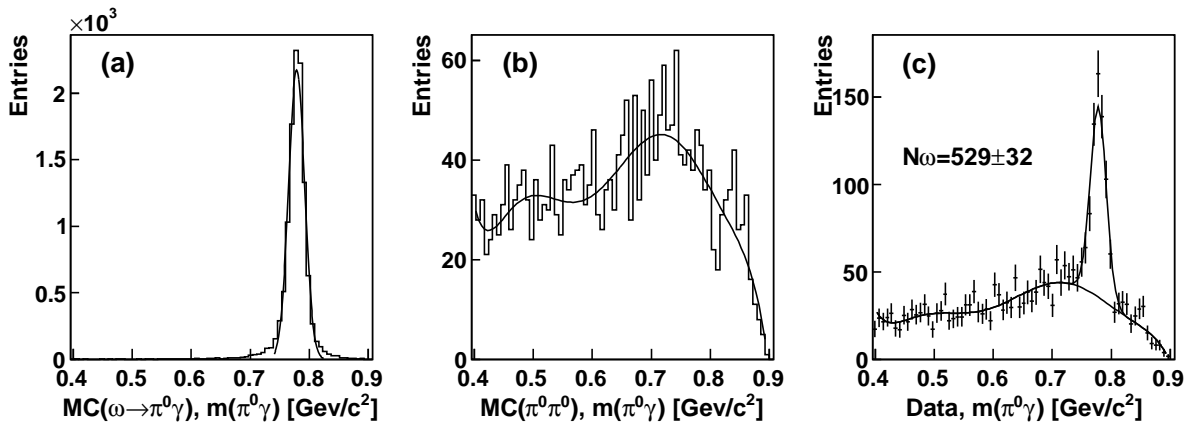}
\caption{
 Same as Fig.~\protect\ref{fig:mpi0g_fit1}, but for $E_\gamma=1325$~MeV.
 A Gaussian fit to the $\omega$ peak results in $\sigma=13.6$~MeV (FWHM=32.0~MeV).
 A polynomial of order eight used to fit the background.
}
 \label{fig:mpi0g_fit3}
\end{figure*}

 The fitting procedure for one energy-angle bin is illustrated in
 Figs.~\ref{fig:mpi0g_fit1}-\ref{fig:mpi0g_fit3},
 showing typical changes in the shape of the $\omega$ peak and the background
 depending on $\theta$ and $E_\gamma$.  
 Panels (a) in these figures depict the $m(\pi^0\gamma)$
 distributions for the MC simulation of $\gamma p\to \omega p\to \pi^0\gamma p$
 fitted with a Gaussian. The choice of the normal distribution for
 the fitting procedure is motivated by the facts that the BW shape of
 the $\omega$ peak is severely cut by phase space near threshold
 and the $m(\pi^0\gamma)$ resolution strongly dominates
 the $\omega$-meson width ($\Gamma=8.49$~MeV~\cite{PDG}),
 especially for the very forward production angles of $\omega$.
 Because the outgoing proton is not detected for those angles,
 it is treated as a missing particle in the kinematic fit,
 resulting in a poorer $m(\pi^0\gamma)$ resolution for the three-cluster events. 
 In the case of the Gaussian fits shown in panels (a), the full width
 at half maximum (FWHM) increases from 19.3~MeV for $E_\gamma=1176$~MeV
 and $\cos\theta=0$ to 29.2~MeV for $\cos\theta=0.93$ at the same energy,
 and to 32.0~MeV for $E_\gamma=1325$~MeV and the same angular bin.

 Panels (b) in these three figures show the $m(\pi^0\gamma)$ distributions
 for the MC simulation of $\gamma p\to \pi^0\pi^0 p$, which is one of
 the background processes. The reaction itself was generated as
 $\gamma p\to \pi^0 \Delta^+(1232)\to \pi^0\pi^0 p$.
 The production angular distribution of $\gamma p\to \pi^0 \Delta$ and
 the rest-frame $\Delta^+(1232)\to \pi^0 p$ decay distribution with respect
 to the $\Delta$'s directions were also generated isotropically.
 A more precise generation of the reaction kinematics were unnecessary
 as these spectra were used only for obtaining initial values for
 parameters of the function describing the experimental background.
 
 To separate the $\omega$ signal from the background,
 the experimental $m(\pi^0\gamma)$ distributions were fitted with the sum of
 a Gaussian, describing the $\omega$ peak, and a polynomial, describing
 the background. These fits are shown in panels (c).
 The order of the polynomial was chosen to be sufficient for a fairly
 good description of the background distribution in the range
 $m(\pi^0\gamma)>0.4$~GeV/$c^2$. Typically, there was no need to use
 a polynomial higher than order eight.
 The initial values for the polynomial coefficients were taken equal
 to the output parameters of the polynomial fit to
 the MC simulation of $\gamma p\to \pi^0\pi^0p$.
 In the fits to the experimental spectra, the centroid and width of
 the Gaussian were fixed to the values obtained from the previous fits
 to the MC simulation for $\gamma p\to \omega p\to \pi^0\gamma p$,
 shown in panels (a). This introduced additional restrictions
 on the background function, improving the separation of the signal
 and background events.
 As also seen in Figs.~\ref{fig:mpi0g_fit1}-\ref{fig:mpi0g_fit3},
 the centroid and width of the Gaussian
 obtained from fitting to the MC simulation of
 $\gamma p\to \omega p\to \pi^0\gamma p$ suit
 the experimental $\omega$ peak very well.
 This confirms the agreement of the energy calibration and
 the detector resolution for the experimental data and the MC simulation.
 
 The number of $\omega \to \pi^0\gamma$ decays
 in the experimental $m(\pi^0\gamma)$ spectra
 was determined from the area under the Gaussian. For consistency,
 the detection efficiency in each energy-angle bin was obtained
 in the same way, i.e., based on a Gaussian fit to the MC simulation for
 $\gamma p\to \omega p\to \pi^0\gamma p$, instead of using the number
 of entries in the $m(\pi^0\gamma)$ spectra.
 This approach allowed one to diminish the impact from some deviation
 of the $\omega$-peak shape from the normal distribution used in the fits.

\section{Experimental results}
  \label{sec:Results}

 The total number of $\omega$ mesons
 produced in each energy-angle bin was obtained by correcting the number
 of $\omega \to \pi^0\gamma$ decays observed
 with the corresponding detection efficiency and
 the $\omega \to \pi^0\gamma$ branching ratio (8.28\%) from the RPP~\cite{PDG}.
 The $\gamma p\to \omega p$ differential cross sections
 were obtained by taking into account the number of protons in the target
 and the photon-beam flux from the tagging facility. Statistical uncertainties
 of the results were calculated from the errors given by the Gaussian fits.
 One contribution to the systematic uncertainty comes from the uncertainty
 in the shape of the background under the $\omega$ peak and
 from the deviation of the $\omega$-peak shape from the normal
 distribution. This uncertainty for one energy-angle bin is independent
 of such uncertainties for other energy-angle bins.
 The magnitude of this uncertainty,
 the average of which comprises 6\% of individual values in
 the differential cross sections, was estimated by fulfilling several
 tests with the experimental and the MC-simulation spectra.
 For the experimental data, the fit was repeated with
 lowering the order of the polynomial used in the fitting procedure
 and changing the $m(\pi^0\gamma)$ range fitted.
 For tests with the MC simulation, the experimental distribution was
 replaced with the sum of the MC simulations for
 $\gamma p\to \omega p\to \pi^0\gamma p$ and $\gamma p\to \pi^0\pi^0p$
 and fitted as experimental data, checking how well the known number
 of $\omega$ events was recovered.
 Another contribution to the systematic uncertainty,
 which is practically the same for all energy-angle bins,
 comes from the determination of the detector acceptance
 and the photon-beam flux; it was estimated as 5\%
 (see Ref.~\cite{etamamic} for more details on this kind
 of systematic uncertainty).
\begin{figure*}
\includegraphics[height=0.97\textwidth, angle=90]{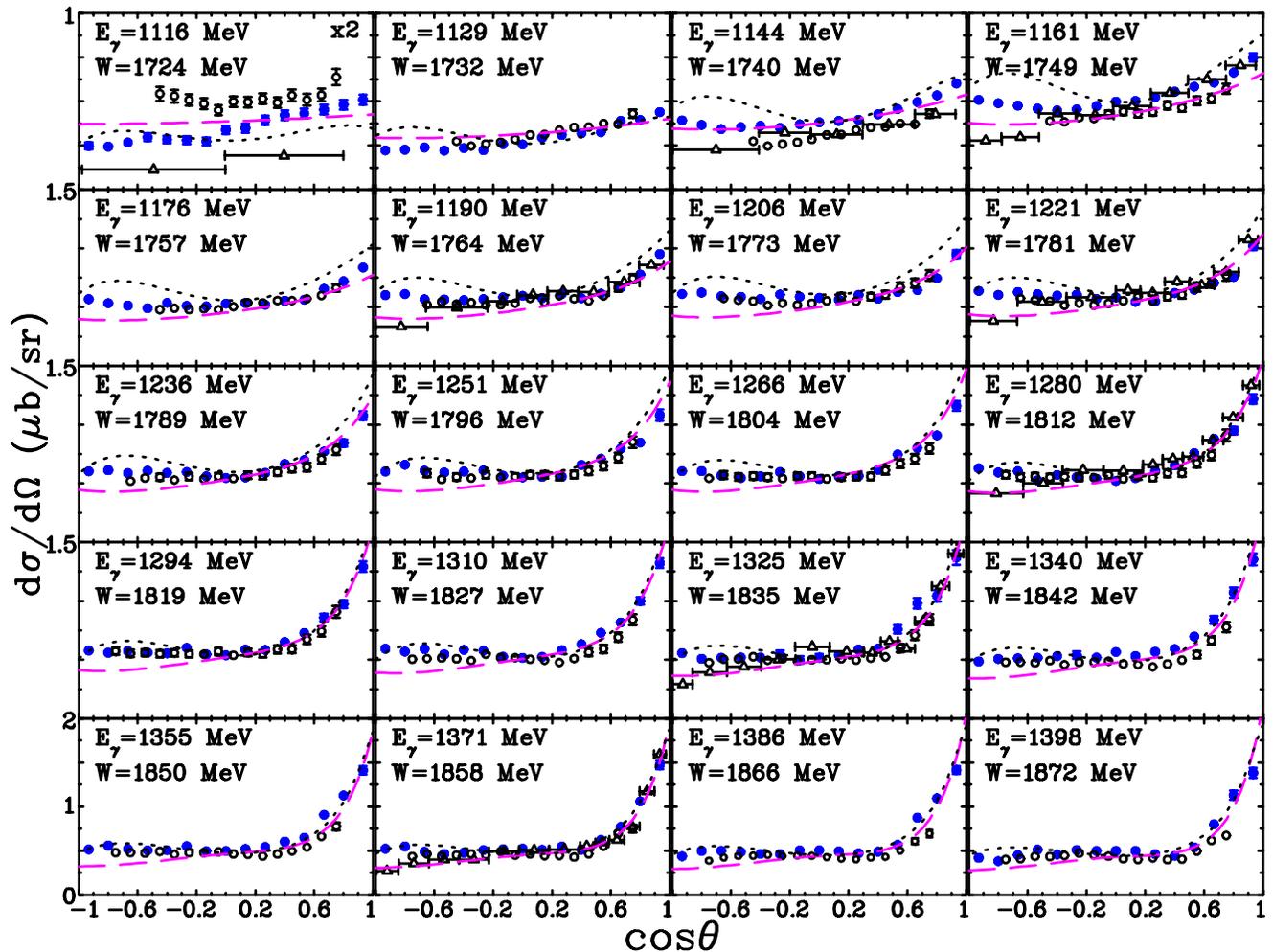}
\caption{
        (Color online) Differential cross sections
	for $\gamma p\to\omega p$ as a function of $\cos\theta$,
        where $\theta$ is the angle between the directions of the $\omega$
        meson and the incident photon in the c.m.\ frame.
        The results of this work are shown by blue dots,
        SAPHIR~\protect\cite{SAPHIR} results by open triangles,
        and CLAS~\protect\cite{CLAS1} results by open circles.
        The vertical error bars of all data points include statistical
        uncertainties only. The horizontal error bars are shown only
        for SAPHIR data, reflecting their nonidentical bins in $\cos\theta$.
        The incident-photon energies of the data points shown for
        previous experiments are within $\pm$5~MeV of $E_\gamma$
        indicated in each panel.
        Model calculations from Ref.~\protect\cite{Beijing}
        are shown by black short-dashed lines and
        from Ref.~\protect\cite{Giessen} by magenta long-dashed lines.
}
 \label{fig:dcs}
\end{figure*}

 In Fig.~\ref{fig:dcs}, the results of this work for the $\gamma p\to \omega p$
 differential cross sections are compared to previous measurements
 at similar energies from SAPHIR~\cite{SAPHIR} and CLAS~\cite{CLAS1}
 and to model calculations from Refs.~\cite{Beijing,Giessen}.
 The vertical error bars shown for all data points in Fig.~\ref{fig:dcs}
 include statistical uncertainties only.
 As seen, the data points of this work cover the full production-angle range and
 have quite small statistical uncertainties. Thus, combining these features
 of the present results with fairly small energy binning (15~MeV in $E_\gamma$)
 makes it possible to study the threshold-region dynamics with much better
 accuracy than before.
 As also seen, the present results are in general agreement with previous
 measurements in the angular range where they overlap. Deviations in absolute
 values at some energies from the results of CLAS, with general agreement
 in the angular dependence at the same time, can be explained
 by a slight difference in $E_\gamma$, which can be very important near threshold.
 The results from SAPHIR for backward angles are smaller than the present and
 other recent measurements.

 The impact of the SAPHIR data on the models can be seen
 in the calculations from the Giessen group~\cite{Giessen}, the predictions
 of which for backward angles follow the behavior of the SAPHIR results.
 The quark-model calculations from Ref.~\cite{Beijing} are in good
 agreement with the present data at higher energies. At lower energies,
 the predictions from Ref.~\cite{Beijing} are larger than experiment,
 especially for backward angles.
 Because the near-threshold amplitudes of this quark model
 are dominated by the under-threshold states $N(1720)3/2^+$ and
 $N(1680)5/2^+$, the comparison with the present data indicates
 that the contributions of those two states could be 
 overestimated in Ref.~\cite{Beijing}. 

 Near-threshold cross sections of good accuracy allow the extraction of
 various useful parameters, including resonance masses as well
 (see, for instance, Refs.~\cite{OPAL,DELPHI,EtaMassA2}).
 In general, the total cross section for an inelastic
 reaction $a~b\to c~d$ with the particle masses $m(a)+m(b)<m(c)+m(d)$
 can be written as $\sigma_t=(q/W)\cdot F(W^2)$, where $W$ is the c.m.\ energy
 and $q$ is the c.m.\ momentum of the final-state particles.
 The factor $F(W^2)$, not vanishing at threshold, comes from the sum of production
 amplitudes squared, and $(q/W)$ from the integration over the final-state phase space.
 Because $W^2$ is linearly related to $E_\gamma$ for meson photoproduction,
 the value $\sigma_t^2$ as a function of $E_\gamma$ reaches zero at
 the threshold energy $E_\gamma=E_{\gamma}^{\mathrm{th}}$ without any
 singularity (i.e., linearly, if the final-state $S$ wave does
 not vanish at threshold).

\begin{figure}
\centering
\includegraphics[height=7.2cm,width=5.8cm,angle=90]{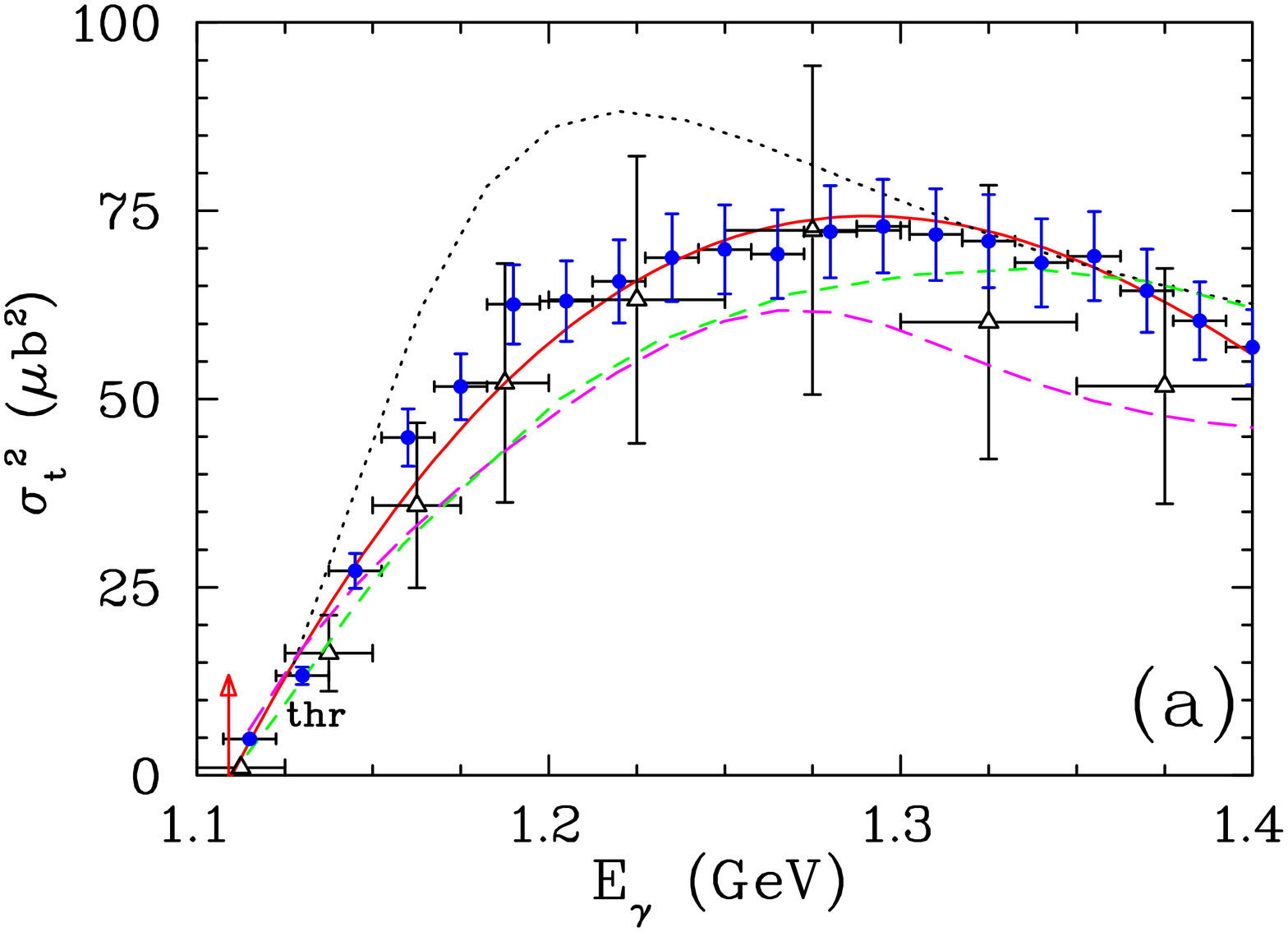}
\includegraphics[height=7.0cm,width=5.8cm,angle=90]{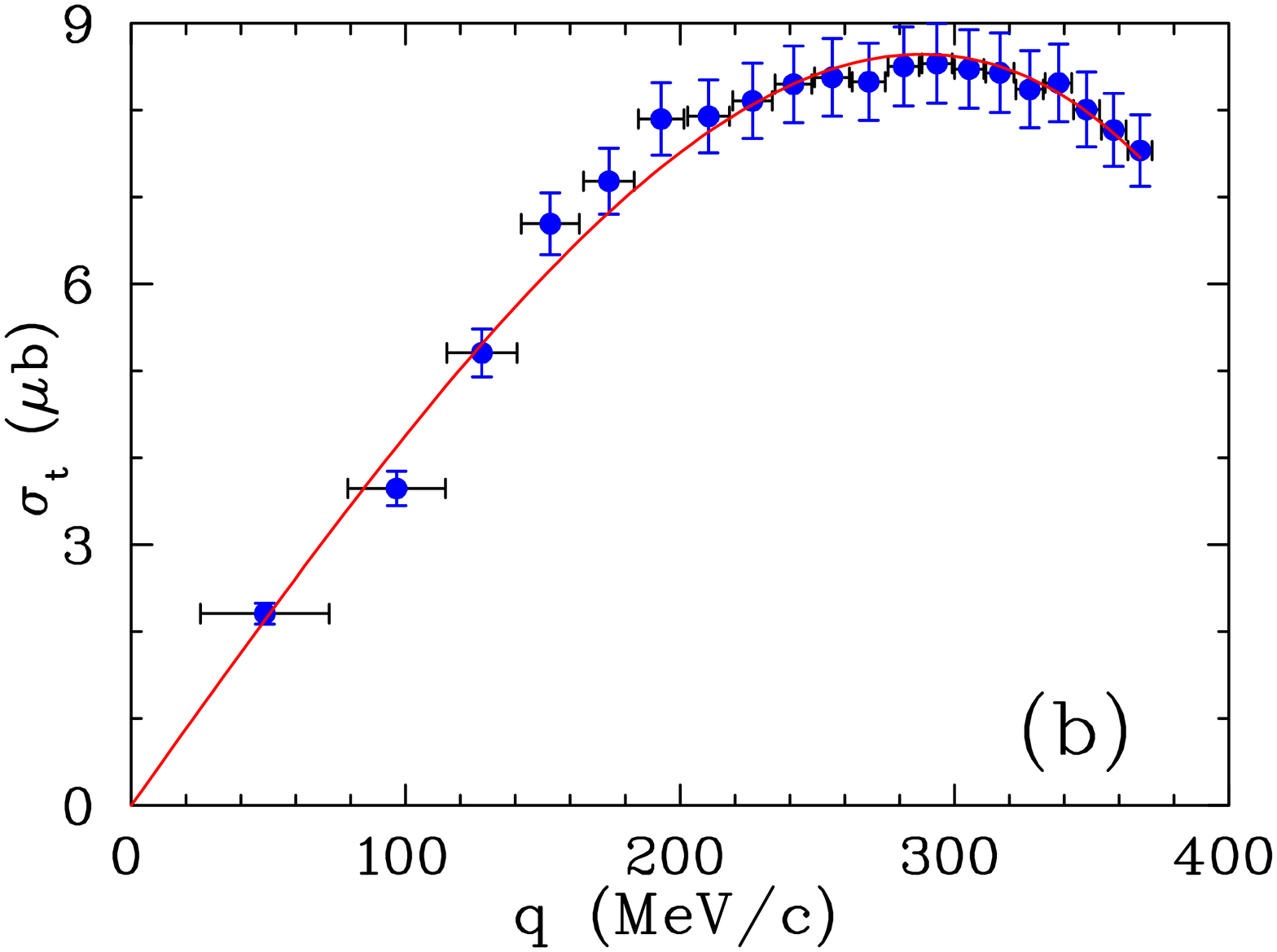}
\caption{(Color online)
  Results of this work (blue dots) for
 the $\gamma p\to\omega p$ total cross sections $\sigma_t$ are shown
 in (a) for $\sigma_t^2$ as a function of the incident-photon
 energy $E_\gamma$ and in (b) for $\sigma_t$ as a function
 of the c.m.\ momentum $q$ of the final-state particles.
 The $\sigma_t$ results from SAPHIR~\protect\cite{SAPHIR}
 are depicted by open triangles. The vertical error bars represent
 the total uncertainties of the results.
 The horizontal error bars reflect the energy binning.
 The red solid line shows the fit of the present data
 (a) with Eq.(\protect\ref{eq1}) and (b) with Eq.(\protect\ref{eq2}).
 The result from the calculation of Ref.~\protect\cite{Beijing} is
 shown by a black short-dashed line, of Ref.~\protect\cite{Giessen}
 by a magenta long-dashed line, and of Ref.~\protect\cite{Paris}
 by a green dashed line.
}
 \label{fig:tcs}
\end{figure}

 The results of this work for $\sigma_t^2(\gamma p\to \omega p)$ are shown
 as a function of $E_\gamma$ in Fig.~\ref{fig:tcs}(a).
 In the same figure, the present results are also compared to model calculations
 from Refs.~\cite{Beijing,Giessen,Paris} and to the results from SAPHIR~\cite{SAPHIR},
 the angular coverage of which (see the horizontal error bars in Fig.~\ref{fig:dcs})
 was almost full, allowing one to extrapolate the differential cross sections
 to the full range.
 The fit of the present $\sigma_t^2$ data with the formula
\begin{equation}
        \sigma_t^2(E_\gamma) =
        b_1 \delta + b_2 \delta^2 + b_3 \delta^3\,,
\label{eq1}
\end{equation}
 with one of four free parameters included in $\delta=E_\gamma-E_{\gamma}^{\mathrm{th}}$,
 is shown in Fig.~\ref{fig:tcs}(a) by a solid red line.
 For the parameter $E_{\gamma}^{\mathrm{th}}$,
 the fit results in the value $(1109.90\pm0.82)$~MeV, corresponding to
 the mass $m_\omega=(783.10\pm0.44)$~MeV/$c^2$.
 It is in good agreement with the RPP value
 $m_\omega=(782.65\pm0.12)$~MeV/$c^2$~\cite{PDG}.
 Although the estimate made here for the $\omega$-meson mass cannot
 compete in precision with the known RPP value, the agreement observed
 indicates the good quality of the present data and the correctness
 of the photon-beam energy calibration, the systematic uncertainty
 in which was determined as 0.5~MeV~\cite{TAGGER2}.

 More traditionally, the $\sigma_t$ behavior of a binary inelastic reaction
 near threshold can be described as a series of odd powers of $q$.
 The results of this work for $\sigma_t(q)$
 are shown in Fig.~\ref{fig:tcs}(b). In the energy range under the study,
 the formula
\begin{equation}
	\sigma_t(q) = a_1 q+ a_3 q^3 + a_5 q^5
\label{eq2}
\end{equation}
 is enough to describe well the present results for $\sigma_t(q)$.
 The fit of these data with Eq.(\ref{eq2}) is shown in Fig.~\ref{fig:tcs}(b)
 by a solid red line, resulting in $a_1 = (4.42\pm 0.14)\cdot10^{-2}~\mu b$/(MeV/$c$),
 $a_3 = -(1.62\pm0.40)\cdot10^{-7}\mu b$/(MeV/$c$)$^3$, and
 $a_5 = -(1.14\pm2.56)\cdot10^{-13}\mu b$/(MeV/$c$)$^5$.
 The linear term is determined here by the $S$ waves only
 (with the total spin 1/2 and/or 3/2),
 while the contributions to the cubic term come
 from both the $P$-wave amplitudes and the $W$ dependence of the $S$-wave
 amplitudes, and the fifth-order term arises from the $D$ waves and
 the $W$ dependences of the $S$ and the $P$ waves.

 The $\sigma_t(\gamma p\to\omega p)$ data near threshold can also be used
 for determining the $\omega N$ scattering length $\alpha_{\omega p}$, defined
 by the threshold relation
 $d\sigma(\omega p\to\omega p)/d\Omega|_{\mathrm{th}} = |\alpha_{\omega p}|^2$
(in reality, it is a combination of two independent $S$-wave
 scattering lengths with total spins 1/2 and 3/2).
 In the VMD framework, $\alpha_{\omega p}$ appears also
 in $\sigma_t(\gamma p\to\omega p)$ near threshold~\cite{Titov}
\begin{equation}
	\sigma_t(\gamma p\to\omega p)|_{\mathrm{th}} =
	\frac{q}{k}\cdot\frac{4\alpha\pi^2}{\gamma^2} \cdot|\alpha_{\omega p}|^2 \,\,,
\label{eqc}
\end{equation}
where $k$ is the c.m.\ momentum of the incident photon at the
 $\gamma p\to\omega p$ threshold, $\alpha$ is the fine-structure
 constant, and $\gamma= 8.53\pm0.14$ is the $\gamma-\omega$
 coupling, as determined from the $\omega\to e^+e^-$ decay width~\cite{PDG}.
 Combining Eq.~(\ref{eqc}) with the $a_1$ value from fitting Eq.~(\ref{eq2})
 to the present $\sigma_t(\gamma p\to\omega p)$ data results in
\begin{equation}
|\alpha_{\omega p}|=\frac{\gamma}{2\pi} \sqrt{\frac{k a_1}{\alpha}}
= (0.82\pm0.03)~\mathrm{fm}~,
\label{alpha}
\end{equation}
which should be considered just as an estimate assuming only the sequence
$\gamma\to\omega,~\omega p\to\omega p$. A more detailed analysis is needed,
however, to exclude contributions from $\gamma\to\rho^0,~\rho^0 p\to\omega p$,
containing in particular $\pi^0$ exchange,
and from a similar transition $\gamma\to\phi$.
Note that the present estimate for $|\alpha_{\omega p}|$ is within the range
defined by other $\alpha_{\omega p}$ values available in the literature:
$(-0.026 + i~0.28)$~fm from the coupled-channel analysis of the $\omega$
production in $\pi$N and $\gamma$N interactions~\cite{Giessen},
$(-0.41 \pm 0.05)$~fm from the QCD sum-rule analysis~\cite{KoHa},
$(1.6 +i~0.30)$~fm from the effective Lagrangian approach based on
chiral symmetry~\cite{Weise}, and
$(-0.44+i~0.20)$~fm from the coupled-channel unitary approach~\cite{Lutz}.
 The dynamical coupled-channel analysis from Ref.~\cite{Paris} yielded
 separate values for two scattering lengths:
 $\alpha_{\omega N}^{1/2}=(-0.0454 -i~0.0695)$~fm and
 $\alpha_{\omega N}^{3/2}=(0.180 -i~0.0597)$~fm,
 which need to be specially combined for comparing with the effective
 scattering length obtained directly from the present data.

 Existing experimental results for the imaginary~\cite{Kotulla}
 and the real~\cite{Friedrich} parts of the $\omega$-nucleus potential
 were recalculated by one of the present authors (V.~M.)
 into the $\omega N$ scattering length, using the connection of
 the meson-nucleus optical potential to the meson-nucleon
 scattering amplitude described in Ref.~\cite{Friedman}
 (details of these calculations are beyond the scope of the present work
 and will be published separately).
 The obtained value, $\alpha_{\omega N} = (-0.17\pm 0.40) + i~(0.79\pm 0.11)$~fm,
 has modulus $|\alpha_{\omega N}| = (0.81\pm 0.41)$~fm that turned out to be
 in agreement, within the uncertainties, with the estimate made in this work. 

 Availability of good-quality data on $\omega$ photoproduction near threshold
 will also allow further analysis of these data for extracting contributions
 from the pion-exchange and the Born nucleon diagrams. The latter diagrams
 contain the coupling vertex $\omega NN$, which determines the $\omega$-exchange
 contribution to $NN$ forces and was extracted earlier in the phenomenological
 analysis of Ref.~\cite{Ted}. Thus, one can check the current understanding
 of the $NN$ potential. Threshold data are important here because,
 at higher energies, the Born contributions decrease and become nonessential
 with respect to the Pomeron-exchange contribution.

\section{Summary and conclusions}
\label{sec:Conclusion}

An experimental study of $\omega$ photoproduction
on the proton was conducted by the A2 Collaboration
at MAMI. The $\gamma p\to\omega p$ differential cross sections
are measured from threshold to $E_\gamma=1.4$~GeV
with 15-MeV binning and full production-angle coverage,
improving significantly the data available for this energy range.
The quality of the present data near threshold gives access
to a variety of physical quantities that can be extracted
by studying the $\omega N$ system.
In particular, our estimate for the $\omega N$ scattering length
is consistent with previous theoretical results, and the estimate
of the $\omega$-meson mass is in good agreement with the RPP~\cite{PDG}
value. The present data are also expected to be invaluable for future
partial-wave and coupled-channel analyses, which could
provide much stronger constraints on the properties of nucleon states
known in this energy range and even reveal new resonances.
A partial-wave analysis with extracting spin-density matrix elements from
the present data is already in progress, being performed by a theoretical
group outside the A2 Collaboration. The results of it will be published
later on separately.

\section*{Acknowledgments}
We thank T.~Barnes, A.~B.~Gridnev, A.~I.~Titov, and Q.~Zhao
for useful remarks and continuous interest in the paper.
The authors wish to acknowledge the excellent
support of the accelerator group and operators of MAMI. This work was
supported by the Deutsche Forschungsgemeinschaft (SFB443, SFB/TR16, and
SFB1044), DFG-RFBR (Grant No. 09-02-91330), the European Community-Research
Infrastructure Activity under the FP6 ``Structuring the European Research Area"
programme (Hadron Physics, Contract No. RII3-CT-2004-506078),
Schweizerischer Nationalfonds, the UK Science and
Technology Facilities Council (STFC 57071/1, 50727/1),
the U.S. Department of Energy (Offices of Science and Nuclear Physics,
 Award Numbers DE-FG02-99-ER41110, DE-FG02-88ER40415, DE-FG02-01-ER41194)
 and National Science Foundation (Grant No. PHY-1039130, IIA-1358175),
 INFN (Italy), and NSERC (Canada). Ya.~I.~Azimov acknowledges support by
the Russian Science Foundation (Grant No. 14-22-00281).
We thank the undergraduate students of Mount Allison University
and The George Washington University for their assistance.


\begin{thebibliography}{99}
\bibitem{CapRob}S.~Capstick and W.~Roberts,
   Prog.\ Part.\ Nucl.\ Phys.\ \textbf{45}, S241 (2000).
\bibitem{PDG} K.~A. Olive  \textit{et al.}, (Particle Data Group),
  Chin.\ Phys.\ C \textbf{38}, 090001 (2014).
\bibitem{ar06}R.~A.~Arndt, W.~J.~Briscoe, I.~I.~Strakovsky, and R.~L.~Workman,
   Phys.\ Rev.\ C\ \textbf{74}, 045205 (2006).
\bibitem{Koniuk} R.~Koniuk and N.~Isgur,
   Phys.\ Rev.\ Lett.\ \textbf{44}, 845 (1980).
\bibitem{VMD}N.~M.~Kroll, T.~D.~Lee, and B.~Zumino,
      Phys.\ Rev.\ \textbf{157}, 1376 (1967);
     J.~J.~Sakurai, Phys.\ Rev.\ Lett.\  \textbf{22}, 981 (1969).
\bibitem{SAPHIR}J.~Barth \textit{et al.},
          	Eur.\ Phys.\ J.\ A\ \textbf{18}, 117 (2003).
\bibitem{Beijing}Q.~Zhao,\ Phys.\ Rev.\ C\ \textbf{63}, 025203 (2001);
  private communication, 2013.
\bibitem{Titov1}Y.~Oh, A.~Titov, and T.~S.~H.~Lee, Phys.\ Rev.\ C\
		\textbf{63}, 025201 (2001).
\bibitem{ba02}H.~Babacan, T.~Babacan, A.~Gokalp, and O.~Yilmaz,
		Eur.\ Phys.\ J.\ A\ \textbf{13}, 355 (2002).
\bibitem{Titov2}A.~I.~Titov and T.~S.~H.~Lee, Phys.\ Rev.\ C\
		\textbf{66}, 015204 (2002).
\bibitem{Giessen}V.~Shklyar, H.~Lenske, U.~Mosel, and G.~Penner,
          	Phys.\ Rev.\ C\ \textbf{71}, 055206 (2005);
                V.~Shklyar, private communication, 2013.
\bibitem{Paris} M.~W.~Paris, Phys.\ Rev.\ C\ \textbf{79}, 025208 (2009).
\bibitem{CLAS1}M.~Williams \textit{et al.},
          	Phys.\ Rev.\ C\ \textbf{80}, 065208 (2009).
\bibitem{CBELSA}A.~Willson, Ph.D. Thesis, Florida State Univ. 2013.
\bibitem{CB}A.~Starostin \textit{et al.}, Phys.\ Rev.\ C\
		\textbf{64}, 055205 (2001).
\bibitem{TAPS} R.~Novotny, 
               IEEE Trans.\ Nucl.\ Sci.\ {\bf 38}, 379 (1991).
\bibitem{TAPS2} A.~R.~Gabler \textit{et al.}, Nucl.\ Instrum.\
                Methods\ A\ \textbf{346}, 168 (1994).
\bibitem{MAMI} H.~Herminghaus {\it et al.},
         IEEE Trans.\ Nucl.\ Sci.\ {\bf 30}, 3274 (1983).
\bibitem{MAMIC} K.-H.~Kaiser \textit{et al.}, Nucl.\ Instrum.\
                Methods\ A\ \textbf{593}, 159 (2008).
\bibitem{TAGGER} I.~Anthony {\it et al.},
  Nucl.\ Instrum.\ Methods\ A\ {\bf 310}, 230 (1991).
\bibitem{TAGGER1} S.~J.~Hall {\it et al.},
  Nucl.\ Instrum.\ Methods\ A\ {\bf 368}, 698 (1996).
\bibitem{TAGGER2} J.~C.~McGeorge {\it et al.},
  Eur.\ Phys.\ J.\ A\ {\bf 37}, 129 (2008).
\bibitem{etamamic} E.~F.~McNicoll \textit{et al.},
                 Phys.\ Rev.\ C\ \textbf{82}, 035208 (2010).
\bibitem{slopemamic} S.~Prakhov {\it et al.},
                 Phys.\ Rev.\ C\ {\bf 79}, 035204 (2009).
\bibitem{OPAL}  K.~Ackerstaff \textit{et al.},
      Phys.\ Lett.\ B \textbf{389}, 416 (1996).
\bibitem{DELPHI} J.~Abdallah \textit{et al.},
      Eur.\ Phys.\ J.\ C\ \textbf{34}, 127 (2004).
\bibitem{EtaMassA2} A.~Nikolaev \textit{et al.},
 		Eur.\ Phys.\ J.\ A\ \textbf{50}, 58 (2014).
\bibitem{Titov}A.~I.~Titov, T.~Nakano, S.~Date, and Y.~Ohashi,
                Phys.\ Rev.\ C\ \textbf{76}, 048202 (2007).
\bibitem{KoHa}Y.~Koike and A.~Hayashigaki, Prog.\ Theor.\ Phys.\
                \textbf{98}, 631 (1997).
\bibitem{Weise}F.~Klingl, T.~Waas, and W.~Weise,
      Nucl.\ Phys.\ A\ \textbf{650}, 299 (1999).
\bibitem{Lutz}M.~F.~M.~Lutz, G.~Wolf, and B.~Friman,
      Nucl.\ Phys.\ A\ \textbf{706}, 431 (2002);
      Erratum, \textbf{765}, 495 (2006).
\bibitem{Kotulla} M.~Kotulla \textit{et al.},
  Phys.\ Rev.\ Lett.\ \textbf{100}, 192302 (2008).
\bibitem{Friedrich} S.~Friedrich \textit{et al.}, 
  Phys.\ Lett.\ B\  \textbf{736}, 26 (2014).
\bibitem{Friedman}E.~Friedman and A.~Gal,
   Phys.\ Rep.\ \textbf{452} 89 (2007).
\bibitem{Ted}C.~Downum, T.~Barnes, J.~R.~Stone, and E.~S.~Swanson,
          	Phys.\ Lett.\ B\ \textbf{638}, 455 (2006).
\end{thebibliography}
\end{document}